\documentclass[a4paper,11pt]{article}
\pdfoutput=1 

\usepackage{jheppub} 

\usepackage[T1]{fontenc} 
\usepackage{xcolor}
\usepackage{makecell}
\usepackage{booktabs}

\newcommand{\MadGraph}{M\protect\scalebox{0.8}{AD}G\protect\scalebox{0.8}{RAPH}}
\newcommand{\MadLoop}{M\protect\scalebox{0.8}{AD}L\protect\scalebox{0.8}{OOP}}

\newcommand{\FastJet}{F\protect\scalebox{0.8}{AST}J\protect\scalebox{0.8}{ET}}
\newcommand{\Sherpa}{S\protect\scalebox{0.8}{herpa}}
\newcommand{\RAMBO}{\texttt{RAMBO}}
\newcommand{\Keras}{\texttt{Keras}}
\newcommand{\TensorFlow}{\texttt{TensorFlow}}
\newcommand{\ONNX}{\texttt{ONNX}}
\newcommand{\CUDA}{\texttt{CUDA}}
\newcommand{\EarlyStopping}{\texttt{EarlyStopping}}
\newcommand{\RatioEarlyStopping}{\texttt{RatioEarlyStopping}}
\newcommand{\ReduceLROnPlateau}{\texttt{ReduceLROnPlateau}}
\DeclareMathOperator{\swish}{swish}
\DeclareMathOperator{\zlog}{zlog}

\title{
    \vspace{-4cm}\hfill {\small IPPP/23/06 \\ \vspace{-0.3cm} \hfill }\vspace{3cm}\\ 
    One-loop matrix element emulation with factorisation awareness}

\author[a]{D. Ma\^{\i}tre,}
\author[a, b, 1]{H. Truong, \note{Corresponding author.}}

\affiliation[a]{Institute for Particle Physics Phenomenology, Durham University, Durham DH1 3LE, UK}
\affiliation[b]{Institute for Data Science, Durham University, Durham DH1 3LE, UK}

\emailAdd{daniel.maitre@durham.ac.uk}
\emailAdd{henry.truong@durham.ac.uk}

\abstract{
    In this article we present an emulation strategy for one-loop matrix elements. This strategy is based on the factorisation properties of 
    matrix elements and is an extension of the work presented in~\cite{Maitre:2021uaa}. We show that a percent-level accuracy can be achieved even 
    for large multiplicity processes. The point accuracy obtained is such that it dwarfs the statistical accuracy of the training sample which allows
    us to use our model to augment the size of the training set by orders of magnitude without additional evaluations of expensive one-loop matrix elements.   
}

\begin{document} 
\maketitle
\flushbottom

\section{Introduction}
\label{sec:intro}

With the high luminosity upgrade of the LHC and the need to investigate high-multiplicity signatures to discover new physics, the need 
to investigate and improve the efficiency of theoretical predictions is gaining momentum~\cite{Azzi:2019yne}. This impetus is further exacerbated by the 
current environment of high energy prices and the immediate need to curb energy consumption in order to fight climate change~\cite{Bloom:2022gux}. 

Multiple approaches have been presented to increase the efficiency of theoretical predictions, addressing a range of different bottlenecks. 
Starting from the calculation of the hard matrix elements, neural networks have been applied to emulate matrix element 
calculations~\cite{Badger:2020uow,Aylett-Bullock:2021hmo,Maitre:2021uaa,Badger:2022hwf,Bishara:2023epi} for tree-level and loop-induced processes. %
Beyond these approaches that aim at replacing the matrix element calculation, work on the matrix element generator itself and PDF evaluation, guided by 
profiling, resulted in large performance improvements~\cite{Bothmann:2022thx}. 

Attention has also been directed to improving the efficiency of phase-space Monte Carlo sampling 
\cite{Bendavid:2017zhk,Klimek:2018mza,Bothmann:2020ywa,Verheyen:2020bjw,Chen_2021,Heimel:2022wyj}, %
to accelerate the simulation of radiation within a jet~\cite{Carrazza:2019cnt,Bothmann:2018trh,Dohi:2020eda}, and  %
to streamline the processes of generating and unweighting simulated event samples %
\cite{Gao:2020zvv,Otten:2019hhl,Hashemi:2019fkn,DiSipio:2019imz,Butter:2019cae,Bishara:2019iwh,Backes:2020vka,Butter:2020qhk,Alanazi:2020klf,Nachman:2020fff,Danziger:2021eeg,Janssen:2023ahv}.

Most of the attention in replacing matrix element calculation with a surrogate has been directed to tree-level or loop induced processes, in this article we shift the focus to %
emulating one-loop matrix elements that are part of a next-to-leading order (NLO) calculation. This was first attempted in Ref.~\cite{Badger:2020uow} using %
a set of neural networks, each focusing on one particular singular region. In this article we adapt the approach developed in Ref.~\cite{Maitre:2021uaa} of %
using known universal singular behaviours of the amplitude as an ansatz for the emulated quantity. The coefficients of that ansatz are learned %
by a neural network (NN) which smoothly interpolates between the singular regions.

The article is organised as follows. Section~\ref{sec:fitting} introduces the factorisation properties and associated functions we will use in our %
ansatz before we detail the construction of the emulator and its training. In Section~\ref{sec:results} we showcase our results. 

Computer code to reproduce the model described in this article is available at \cite{fame_repo}.

\section{Fitting framework for one-loop matrix element}\label{sec:fitting}

For this work we consider the emulation of one-loop 
$e^+e^-\rightarrow Z / \gamma^{*} \rightarrow q\bar q +n_{g} g$
matrix elements for $n_{g}$ up to and including 3, which corresponds
to events with up to 5 particles in the final-state. We denote the number of final-state
partons as $n$. Instead of emulating the matrix element itself we build
a surrogate for the related so-called k-factor
\begin{equation}
    k_{n} = \dfrac{2 \Re \left\{ \mathcal{M}^{(n, 0)} \mathcal{M}^{(n, 1)\, *} \right\} }{ |\mathcal{M}^{(n, 0)}|^{2} } \equiv \dfrac{|\mathcal{M}^{(n, 1)}|^{2}}{|\mathcal{M}^{(n, 0)}|^{2}} \, ,
    \label{eq:k_factor}
\end{equation}
where $\mathcal{M}^{(n, \ell)}$ denotes the amplitude
for a process with $n$ final-state partons, at loop-order
$\ell$. We will refer to the interference term in the numerator
as the one-loop matrix element henceforth and introduce
this notation for brevity.
The sum/averaging over colour and helicity is implicit for
both the one-loop matrix element and the tree-level
matrix element. The numerator in (\ref{eq:k_factor}) is the
finite part\footnote{See Section 3.2 and Appendix A.1 in Ref.~\cite{Hirschi:2011pa} for explicit definitions.} of the interference between the tree-level and 
one-loop level matrix element, where the conventional
dimensional regularisation (CDR) scheme is used.

We choose to emulate the k-factor instead of the one-loop matrix
element directly because the division of the tree-level matrix
element normalises the infrared divergences occurring for soft
and collinear external particles. However, there are still
logarithmic divergences that remain from the loop integral. Another 
advantage is that the scale of the k-factors is naturally of
the order unity, making it more amenable for emulation.

In the following, we describe how we apply the same approach as the
factorisation-aware formalism introduced in Ref.~\cite{Maitre:2021uaa}
to encapsulate the more complex structure of the one-loop
matrix element to construct an accurate emulator for the
one-loop k-factors that is robust against single collinear or soft divergences.

An additional complication with one-loop matrix elements
is that they are evaluated at a given renormalisation scale.
This dependence can be derived from first principle, but we choose
to instead incorporate this dependence into our neural network emulator
as an input.
This method, so-called parametric
neural networks, has been utilised in other contexts
\cite{Baldi:2016fzo,Ghosh:2021roe}.

\subsection{Antenna functions}\label{sec:antenna}
In building an emulator for tree-level matrix elements,
Catani-Seymour dipoles \cite{Catani:1996vz} are sufficient
to explain all single divergences arising in phase-space. For
one-loop matrix elements we utilise antenna functions
which fulfil a similar purpose of providing a set of functional
behaviours to build the matrix element out of.

Antenna functions as given in Ref.~\cite{Gehrmann-DeRidder:2005btv}
are derived from physical colour-ordered matrix elements and
by construction have the correct infrared behaviour when specific sets of
particles become unresolved. For our purposes we require the set of 
antenna functions describing the scenario of one particle
becoming unresolved at tree-level and one-loop level, namely,
following the notation of Ref.~\cite{Gehrmann-DeRidder:2005btv},
these are the leading colour three-parton antenna functions
$X_{ijk}^{0}$ and $X_{ijk}^{1}$, respectively. $X_{ijk}^{0}$
describes all configurations where parton $j$ becomes unresolved,
where $i$ and $k$ are the hard partons. $X_{ijk}^{1}$ correctly
reproduces the single soft and collinear singularity structure
appearing in the one-loop singular functions \cite{Bern:1999ry}.
From this description it is clear that the antenna functions
depend on the momenta in the full $(n+1)$-body phase-space.

In one antenna function, there are two hard partons
which can both radiate off one unresolved parton. This is in
contrast to the dipole function which only has an unresolved
parton emitting from one parton. In that sense, a single antenna
function is a linear combination of two dipole functions where
the emitter and spectator are swapped. The advantage of this is
that there are generally fewer antenna functions to consider,
especially when the multiplicity increases.

Although the one-loop matrix elements that we use for fitting are not
colour-ordered, the antenna functions nevertheless provide
a set of useful functions that allow the neural network emulator
to form accurate approximations of the one-loop k-factor.

Since we are emulating the finite part of the one-loop matrix
element we need to take care to extract all the finite parts
from the one-loop antenna functions $X_{ijk}^{1}$. The
full expression for the one-loop antenna we use is given as
\begin{equation}
    X_{ijk}^{1,F}  = \mathcal{F}inite(X_{ijk}^{1}) + \dfrac{11}{6}\log\left(\dfrac{\mu_{R}^{2}}{s_{ijk}} \right) X_{ijk}^{0} + \mathcal{F}(\mathbf{I}_{ij}^{(1)}(\epsilon, s_{ij})) X_{ijk}^{0} \, .
    \label{eq:finite_X31}
\end{equation}
where the superscript $F$ in $X_{ijk}^{1,F}$ denotes the
one-loop antenna function with all finite parts extracted\footnote{Henceforth
we refer to $X_{ijk}^{1,F}$ as the one-loop antenna function,
unless explicitly state otherwise.}.
Most of the finite parts of the antenna function are extracted
in the term $\mathcal{F}inite(X_{ijk}^{1})$ as given
in Ref.~\cite{Gehrmann-DeRidder:2005btv}. The second term
adjusts the renormalisation scale of the antenna function from
the invariant mass of the antenna partons, $s_{ijk} = s_{ij} + s_{ik} + s_{jk}$,
to the renormalisation scale the one-loop matrix element is
evaluated at, $\mu_{R}^{2}$. The final term extracts the
remaining finite parts from the infrared singularity operators,
where their explicit expressions are given in Appendix~\ref{appendix:finite_terms}.

\subsection{Factorisation of matrix elements}\label{sec:me_factorisation}
In the following, we describe the factorisation of matrix
elements in the language of antenna functions \cite{Gehrmann-DeRidder:2005btv}.

\begin{figure}
    \includegraphics[width=\linewidth]{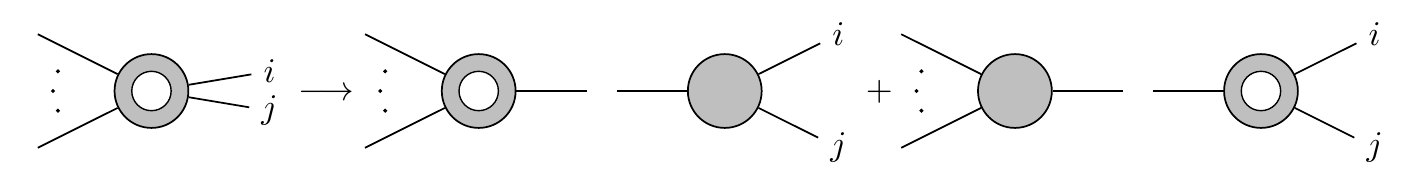}
    \caption{Diagram illustrating factorisation of one-loop
    matrix element.
    In a singly unresolved limit, the $(n+1)$-body one-loop matrix
    element tends to the sum of a $n$-body one-loop matrix element
    multiplied by a tree-level splitting kernel and a $n$-body tree-level
    matrix element multiplied by a one-loop splitting kernel.
    The tree-level elements are drawn as a fully filled in
    circle, while the one-loop elements are drawn as a donut shape.}
    \label{fig:ol_factorisation}
\end{figure}

Tree-level matrix elements in $(n+1)$-body phase-space
can be factorised in the single soft and collinear limits as
\begin{equation}
    |\mathcal{M}^{(n+1, 0)}|^{2} \longrightarrow X_{ijk}^{0} |\mathcal{M}^{(n, 0)}|^{2} \, ,
    \label{eq:tree_factorisation}
\end{equation}
where $|\mathcal{M}^{(n, 0)}|^{2}$ is the reduced matrix element
in $n$-body phase-space and $X_{ijk}^{0}$ is the three-parton tree-level
antenna function introduced in Section~\ref{sec:antenna}.
The one-loop matrix element similarly
exhibits factorisation in the soft and collinear limits.
This has been extensively studied \cite{Bern:1994zx,Bern:1995ix,Bern:1999ry,Kosower:1999xi,Kosower:1997zr}
with the splitting kernels computed \cite{Kosower:1997zr,Kosower:2002su,Gehrmann-DeRidder:2005btv}.
Schematically, in the single soft and collinear limits, the one-loop
matrix element can be deconstructed into
\begin{equation}
    |\mathcal{M}^{(n+1, 1)}|^{2} \longrightarrow X_{ijk}^{0} |\mathcal{M}^{(n, 1)}|^{2} + X_{ijk}^{1,F} |\mathcal{M}^{(n, 0)}|^{2} \, ,
    \label{eq:ol_factorisation}
\end{equation}
where $X_{ijk}^{1,F}$ is the three-parton one-loop antenna function.
This equation can be thought of as a tree-level splitting
kernel multiplied by a one-loop reduced matrix element,
plus a one-loop splitting kernel multiplied by a tree-level
reduced matrix element. This is illustrated
pictorially in Figure~\ref{fig:ol_factorisation}.

\subsection{Ansatz for the k-factor}\label{sec:ansatz}
Given that both the tree-level and one-loop level matrix
element factorise in the soft and collinear limits,
we can rewrite the $(n+1)$-body k-factor in these limits as
\begin{align}
    k_{n+1} &\longrightarrow \dfrac{X_{ijk}^{0} |\mathcal{M}^{(n, 1)}|^{2} + X_{ijk}^{1,F} |\mathcal{M}^{(n, 0)}|^{2}}{X_{ijk}^{0} |\mathcal{M}^{(n, 0)}|^{2}} \nonumber \\
    k_{n+1} &\longrightarrow \dfrac{|\mathcal{M}^{(n, 1)}|^{2}}{|\mathcal{M}^{(n, 0)}|^{2}} + \dfrac{X_{ijk}^{1,F}}{X_{ijk}^{0}} \\
    k_{n+1} &\longrightarrow k_{n} + \dfrac{X_{ijk}^{1,F}}{X_{ijk}^{0}} \nonumber \, .
    \label{eq:k_factorisation}
\end{align}
where we see that the k-factor tends to a sum of the reduced k-factor, $k_{n}$, and a ratio
of antenna functions. By summing over ratios of antenna functions for all limits of
a given process, we can construct an ansatz for the k-factor over all of phase-space.
This informs our ansatz for the k-factor, which can be given as
\begin{equation}
    k_{n+1} = C_{0} + \sum_{\{ijk\}} C_{ijk}\dfrac{X_{ijk}^{1,F}}{X_{ijk}^{0}} \,
    \label{eq:ansatz}
\end{equation}
where $C_{0}$ and $C_{ijk}$ are coefficients fitted by the neural network.
$C_{0}$ is an additive term aiming to model the reduced
k-factor, and $C_{ijk}$ are multiplying the ratio of
antenna functions to fit the single collinear, and
soft limits in multiple regions of phase-space.
The sum over $\{ijk\}$ denotes the sum over
the relevant permutations of final-state partons.
This sum allows the neural network to make use of all the
provided antenna functions to make
an approximation of the colour-summed matrix element. Outside of
the soft and collinear limits, the ansatz makes use of the excellent
interpolation abilities of neural networks to fit the k-factor in
the well-behaved regions of phase-space. This is possible because
the antenna functions are not singular outside of these limits.
The full set of antenna functions which we implement into our
model is given in Appendix~\ref{appendix:antenna_functions}.

Since the k-factor has infrared divergences arising from
unresolved partons in the final-state being removed, and with
the appropriate antenna functions accounting for the
logarithmic divergences from the loop momenta, the
challenging task of fitting a rapidly varying function
over phase-space is reduced to fitting a group of
well-behaved coefficients that dictate how to suitably utilise
the antenna functions.

\subsection{Building the neural network emulator}\label{sec:nn}
\subsubsection*{Dataset generation}
Phase-space is sampled uniformly using the {\RAMBO} algorithm
\cite{Kleiss:1985gy} with a centre-of-mass energy
$\sqrt{s_{\mathrm{com}}} = 1000$ GeV. The global phase-space
generation cut is set to $y_{\mathrm{cut}} = 0.0001$.
We have shown in Ref.~\cite{Maitre:2021uaa} that the accuracy of the emulation
is not greatly affected by the generation cut but that the extrapolation
performance is increased with a more inclusive cut, so we have chosen this
value. 
{\FastJet} \cite{Cacciari:2011ma,noel_dawe_2021_4446849} is
used to exclusively ($d_{\mathrm{cut}} = 0.01 \times s_{\mathrm{com}}$)
cluster final-state jets with the $e^{+}e^{-}$ $k_{t}$
algorithm such that there is at most a single unresolved parton.

For each phase-space point generated, we sample a
renormalisation scale, $\log(\mu_{R})$, from a uniform distribution
with end points at $[\log(\sqrt{s_{\mathrm{com}}}/4), \log(4\sqrt{s_{\mathrm{com}}})]$.
In other words, we sample the renormalisation scale logarithmically.
We observe that the neural network manages to learn the
renormalisation scale dependence well, therefore opt to
sample $\mu_{R}$ in a wider range than is usually used
for the conventional scale variations which varies $\mu_{R}$
up and down by factors of 2.

Generated phase-space points are fed to {\MadGraph} \cite{Alwall:2014hca,Hirschi:2011pa}
to compute the tree-level and one-loop level matrix elements (using {\MadLoop}).
For each phase-space point, we use the corresponding
renormalisation scale sampled and evaluate the strong coupling
constant at this scale using the NNPDF-4.0 NNLO PDF set
\cite{NNPDF:2021njg} with the LHAPDF6 interface \cite{Buckley:2014ana}.
All external particles are treated as massless and $m_{Z} = 91.188$ GeV.

We generate 1100k data points in total, using 100k points for
training and validation (80:20 split), leaving 1 million points
for independent evaluation of model performance.
Training on a limited dataset is a realistic scenario
for processes which are prohibitively expensive, and we show
that it is possible to build an accurate emulator with the relatively
small number of data points. Note that because we are
sampling $\mu_{R}$ along with phase-space simultaneously,
there is an extra dimension in the sampled space. This means that the 100k
training points we have are not comparable to 100k training points
if we had not sampled over $\mu_{R}$. In practice, we have found that
sampling over $\mu_{R}$ has a small impact on accuracy but
opt to go this route to have the flexibility to predict over
a range of $\mu_{R}$.

\subsubsection*{Inputs to model}
As inputs to the neural network we provide the 4-momenta of all
final-state partons. The renormalisation scale enters the network
as $\log(\mu_{R})$ as we expect the dependence on $\mu_{R}$
to be in the form of a logarithm. Following Ref.~\cite{Maitre:2021uaa}
we include the phase-space mapping variables to aid the network in learning
the reduced matrix element information. More specifically, we include as
inputs $r$ and $\rho$ from Ref.~\cite{Kosower:1997zr}
\begin{align}
    r_{ijk} &= \dfrac{s_{jk}}{s_{ij} + s_{jk}} \\
    \rho_{ijk} &= \sqrt{1 + \dfrac{4r(1-r)s_{ij}s_{jk}}{s_{ijk}s_{ik}}} \, ,
    \label{eq:map_variables}
\end{align}
where $i$ and $k$ are the hard radiating partons, and $j$ is
the unresolved parton. We include the subscript
$_{ijk}$ on $r$ and $\rho$ to represent the explicit
dependence on the specific set of momenta used to calculate them.
To improve training, we transform these variables as
$r \rightarrow \log(r)$ and $\rho \rightarrow \log(\rho - (1 - \varepsilon))$,
where $\varepsilon$ is a small constant added to improve the
numerical stability. It is taken to be $\varepsilon = 10^{-8}$.
An additional input that we have observed to increase
accuracy of the emulator are the Mandelstam kinematic invariants.
These are fed into the model as $\log(s_{ij})$ for all
pairs of final-state particles. All inputs are
standardised to zero mean and unit variance.

\begin{figure}[t]
    \includegraphics[width=\linewidth]{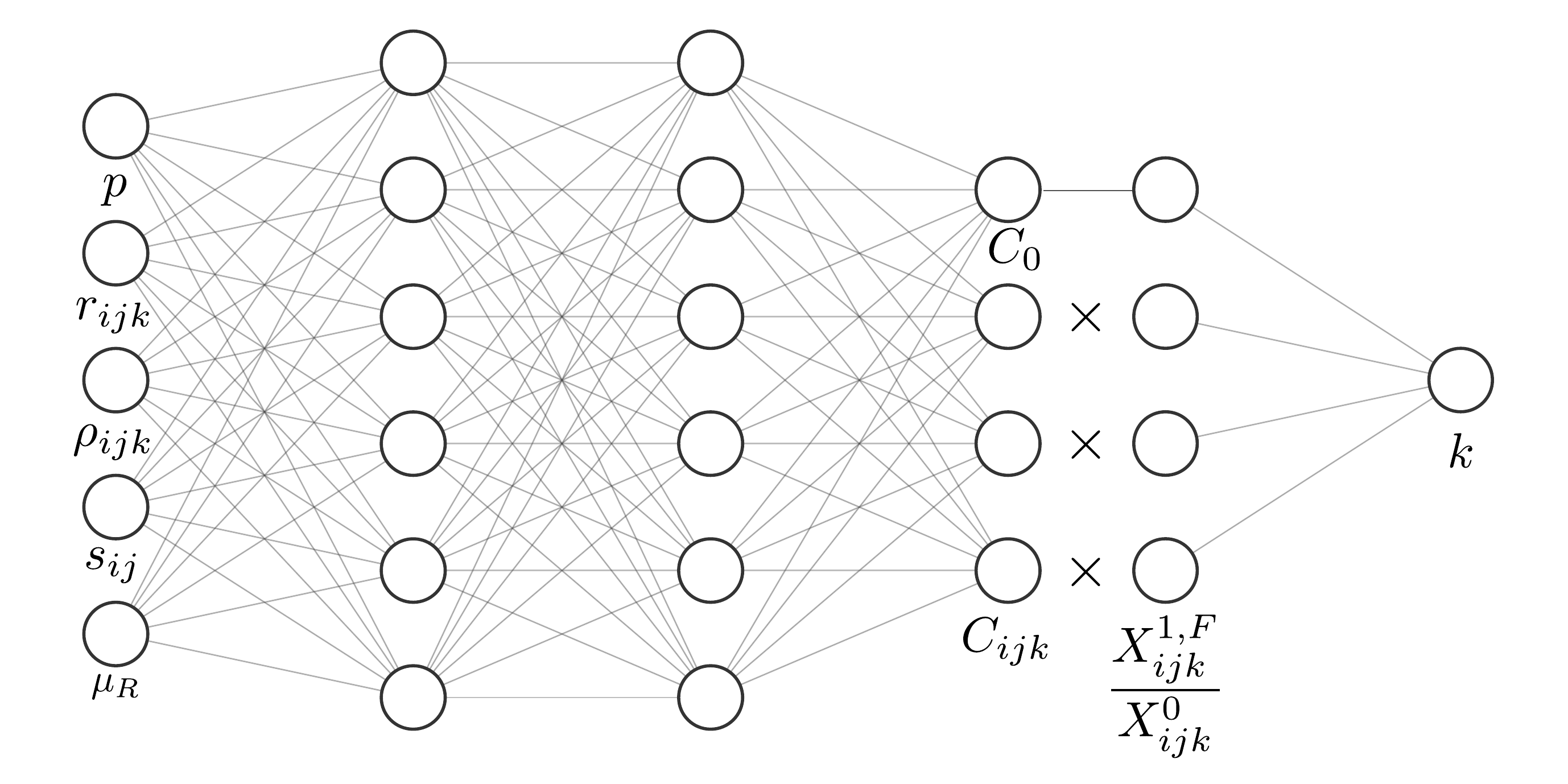}
    \caption{A schematic diagram of the neural network emulator.
    The emulator is a dense neural network with inputs:
    phase-space points, $p$, momenta mapping variables,
    $r_{ijk}$ and $\rho_{ijk}$, and kinematic invariants, $s_{ij}$.
    The outputs of the network are the fitted coefficients, $C_{0}$ and $C_{ijk}$,
    as given in (\ref{eq:ansatz}).}
    \label{fig:nn}
\end{figure}

\subsubsection*{Outputs of model}
The outputs of the neural network are the fitted coefficients $C_{0}$ and $C_{ijk}$
in (\ref{eq:ansatz}), which when combined with the antenna functions
produces an approximation of the k-factor. We then recover the
one-loop matrix element by multiplying by the corresponding
tree-level matrix element. We do not provide the tree-level matrix
element in the emulator as the evaluation time is generally
much lower than that of the one-loop matrix element, and it is usual 
for one-loop matrix elements to be evaluated at phase-space point sets
where the tree-level matrix element has already been unweighted, so that 
only the k-factor is required.

To train the network we we compare the target k-factors from {\MadGraph} to the
predictions from the neural network in the loss function by providing the network
with the antenna functions. 
The target distribution is standardised to zero
mean and unit variance. Since the k-factors are of order unity
we do not need to do any additional pre-processing to aid the network
in training.

\subsubsection*{Neural network architecture}
A schematic of the neural network model is given in Fig.~\ref{fig:nn}.
We build the neural network emulator with {\Keras}
\cite{chollet2015keras} and {\TensorFlow} \cite{tensorflow2015-whitepaper}.
The emulator is a dense neural network with
three hidden layers of 64 nodes each. This network
size was chosen with consideration given to the number
of training samples and to reduce the discrepancy between training
and validation loss
(i.e. a larger network is more prone to overfitting on the
training set if there is insufficient data to fit the additional weights).
For the remaining hyperparameters we summarise the network
architecture in Table~\ref{table:parameters}.
\begin{table}
    \caption{Hyperparameters of the neural network and their values.}
    \begin{center}
        \begin{tabular}{ll}
            \hline
            Parameter & Value \\
            \hline
            Hidden layers & 3 \\
            Nodes in hidden layers & [64, 64, 64] \\
            Activation function & $\swish$ \cite{2017arXiv170203118E,2017arXiv171005941R}\\
            Weight initialiser & Glorot uniform \cite{Glorot10understandingthe} \\
            Loss function & MAE (k-factor), MSE (one-loop matrix element) \\
            Batch size & 256 \\
            Optimiser & Adam \cite{kingma2017adam} \\
            Learning rate & $10^{-3}$ \\
            Callbacks & {\EarlyStopping}, {\RatioEarlyStopping}, {\ReduceLROnPlateau} \\
            \hline
        \end{tabular}
    \end{center}
    \label{table:parameters}
\end{table}

We choose to use the mean absolute error (MAE) as the loss
function for training because it is precisely the error measure we
would like to minimise.
The error for one prediction is given as
\begin{equation}
    k_{\mathrm{true}} - k_{\mathrm{pred}} = \dfrac{|\mathcal{M}^{(n, 1)}|^{2}_{\mathrm{true}} - |\mathcal{M}^{(n, 1)}|^{2}_{\mathrm{pred}}}{|\mathcal{M}^{(n, 0)}|^{2}_{\mathrm{true}}} = \Delta \, ,
    \label{eq:mae}
\end{equation}
where the error in the one-loop matrix element normalised
by the tree-level matrix element is what we want the
neural network to minimise.
Since k-factors are a ratio of matrix elements, the numerical
values it can take are not unique for a given value of the
tree-level matrix element and/or one-loop matrix element.
For example, for two similar values of the k-factor, the scales
of the matrix elements going into each ratio may be vastly different.
To ensure that the network remains accurate for large values
of the tree-level matrix element, where corresponding corrections
contribute more to the total cross-section, we weight the training
points by
\begin{equation}
    w_{i} = \log\left( \dfrac{|\mathcal{M}^{(n, 0)}|^{2}_{i}}{\min(|\mathcal{M}^{(n, 0)}|^{2}_{i})} \right) \, ,
\end{equation}
where the $i$ index denotes individual training samples.
Although we use the MAE on the k-factors as the training loss,
we terminate model training based on the one-loop matrix
element accuracy. This is done by monitoring the mean squared
error (MSE) between the model prediction and corresponding truth value
at the end of each training epoch.
This takes advantage of the compact k-factor distributions
for training purposes, but bases model selection on the
performance for the physical one-loop matrix elements.

To reduce the effects of overfitting we have two {\EarlyStopping}
criteria: the first is to stop training once the validation loss has
not improved in 100 epochs, and the second is a {\RatioEarlyStopping}
which terminates training if the ratio of training loss to validation
loss drops below a certain threshold. We take this threshold
to be $0.9$. We also use the {\ReduceLROnPlateau}
callback as a way to adapt the learning rate during training.
The learning rate is reduced by a factor of 0.7 whenever
validation loss plateaus with a patience of 20 epochs. We find
that with these hyperparameters we achieve a balance of reducing
overfitting and quick training times. On average the models
train in approximately 20 minutes on an Nvidia P100 GPU.

Although we build and train our model using {\TensorFlow},
we deploy the model using the Open Neural Network Exchange ({\ONNX})
runtime \cite{onnxruntime}
with the {\CUDA} execution provider to run predictions on an
Nvidia P100 GPU. With the optimised operations in the {\ONNX}
runtime, we see that compared to {\TensorFlow} the model inference
time is reduced by an order of magnitude or more. Another advantage
is that it provides flexibility to move the pipeline
away from {\TensorFlow} on $\mathtt{Python}$ to a more generic interface
to the neural network model. One example would be to integrate
the {\ONNX} model into a $\texttt{C++}$ workflow for use with current event
generators to replace the one-loop provider with a neural
network emulator. This workflow was recently seen for tree-level
matrix element emulation within the {\Sherpa} framework in the context
of accelerating the generation of unweighted events \cite{Janssen:2023ahv}.

In addition to the {\CUDA} execution provider, we will use
the {\ONNX} runtime CPU execution provider to compare with
{\MadGraph} for a comparison of single CPU core performance.
This will be the closest to a real world benchmark as event
generators typically generate events on a single core.

Although we find that the neural network models converge well,
to account for stochasticity in the training, the random
initialisation of model parameters, and to reduce
variance on predictions, we initialise 20 models for training
and use the mean of these models as our model prediction. This
ensembling will also give a measure of the uncertainty
due to the neural network optimisation, by using the standard
deviation of the 20 replica model predictions.

\begin{figure}
    \centering
    \includegraphics[scale=0.57]{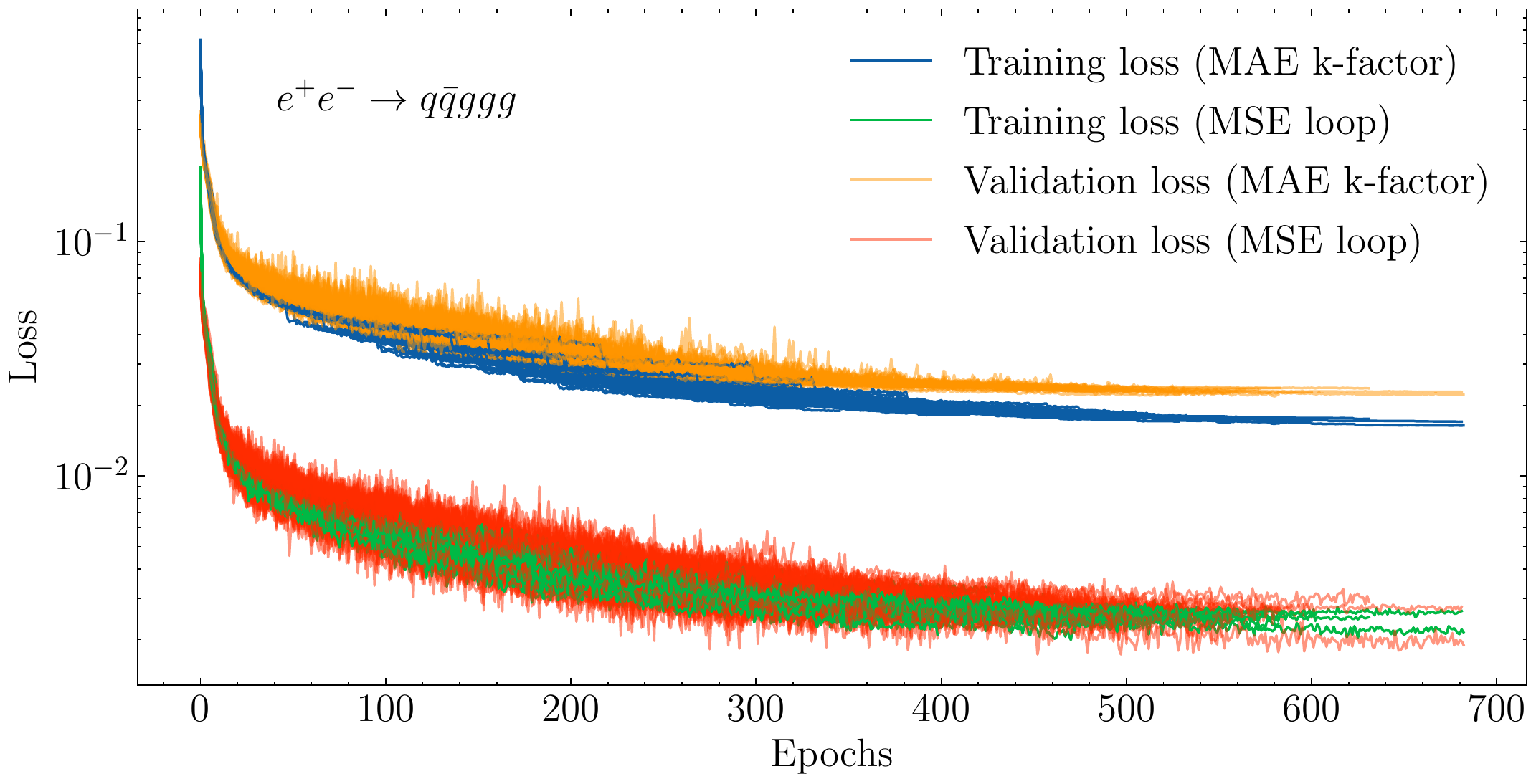}
    \caption{Training and validation losses of the 20 independent
    initialisations of the ensemble replicas plotted as individual
    curves for the 5 parton model. 
    We plot the MAE loss for the k-factor and the MSE loss for
    the one-loop matrix element. The scale difference between the MAE
    and MSE losses is a consequence of the form of the loss functions and
    is not surprising. We see that the replica models all
    converge to similar point with the validation loss being close to
    the training loss when training is terminated. The step feature in
    the training loss is due to the {\ReduceLROnPlateau} callback.}
    \label{fig:loss_curves}
\end{figure}

We plot in Figure~\ref{fig:loss_curves} the losses of the 20 replica
models for the 5 jet process, where we have plotted the loss for the
k-factor (MAE) and the one-loop matrix element (MSE).
We can see that the models have all converged to a similar point
when training is terminated. The noise in the validation loss at
the beginning of training can be attributed to the fitting of the
coefficients: when the model is learning how to pick the relevant
combination of antenna functions there can be large variations in
the prediction, however, the variations become much smaller once
the model learns the factorisation properties and converges. We
can see the variations in the validation loss are small at the
end of training, and that there is not a large discrepancy between
the training and validation losses, as enforced by the {\RatioEarlyStopping}
callback. Furthermore, since the training of the network is dictated
by the MAE loss, the more apparent noise in the MSE loss was foreseeable
as the losses are unlikely to respond the same way to updates of
the model weights.

\section{Results}\label{sec:results}
\begin{figure}
    \centering
    \includegraphics[width=\linewidth]{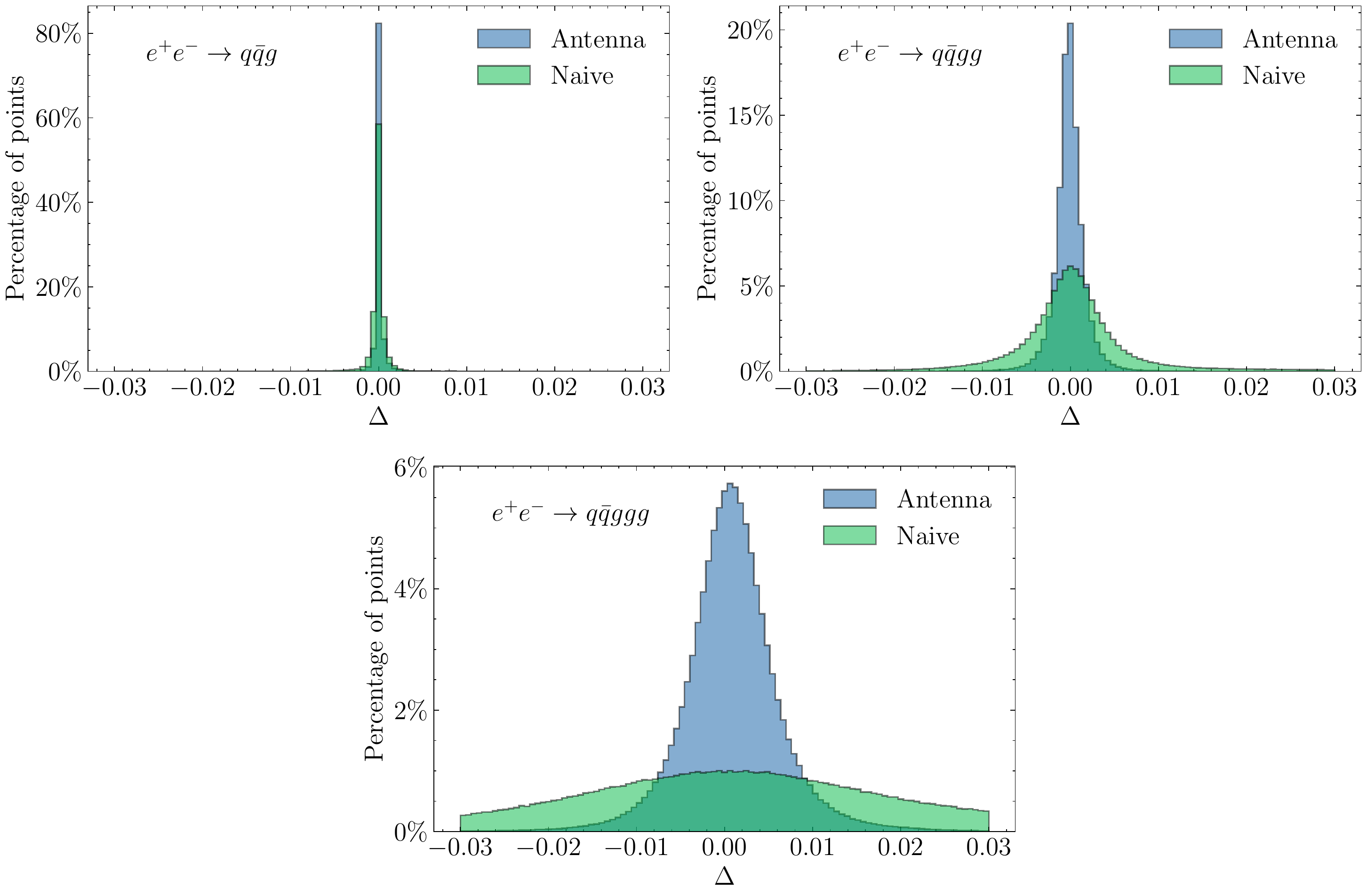}
    \caption{Error distribution in terms of $\Delta$ for all multiplicities. `Antenna'
    model is as described in this article, and `naive' model is a simple densely connected
    neural network model without any factorisation properties built in.
    We keep the horizontal axis scale fixed for all subplots to make it easier to compare
    accuracy across the different multiplicities.}
    \label{fig:delta_distribution}
\end{figure}

In this section we present results for our NLO QCD k-factor
emulator for $e^{+}e^{-}$ annihilation into up to 5-jets.
First we show a comparison between our model described in
Section~\ref{sec:nn} which we label `antenna',
and a `naive' model with no factorisation properties built
into the emulator: a densely connected neural network
with the parameters given in Table~\ref{table:parameters}
that directly predicts the k-factor.
For the `naive' models we train with the MAE on the k-factors
as the loss function with no modifications and terminate training
based on this loss. As with the `antenna' models, we also
ensemble 20 individual replica models for the `naive' model predictions.
In Figure~\ref{fig:delta_distribution} we compare histograms
of $\Delta$ for all final-state multiplicities between these two models.
It is immediately clear that building
in the factorisation structure of the matrix elements greatly increases
accuracy, with increasing relative improvements for the higher multiplicity cases.
The `antenna' error distributions are symmetric, strongly peaked around
the ideal value of 0, and with tails falling off rapidly. We
see the general trend of increasing multiplicity decreases accuracy,
however we observe that the bulk of the 5 jet final-state is
within percent accuracy and with the lower multiplicities well
below this.

\begin{figure}
    \centering
    \includegraphics[scale=0.6]{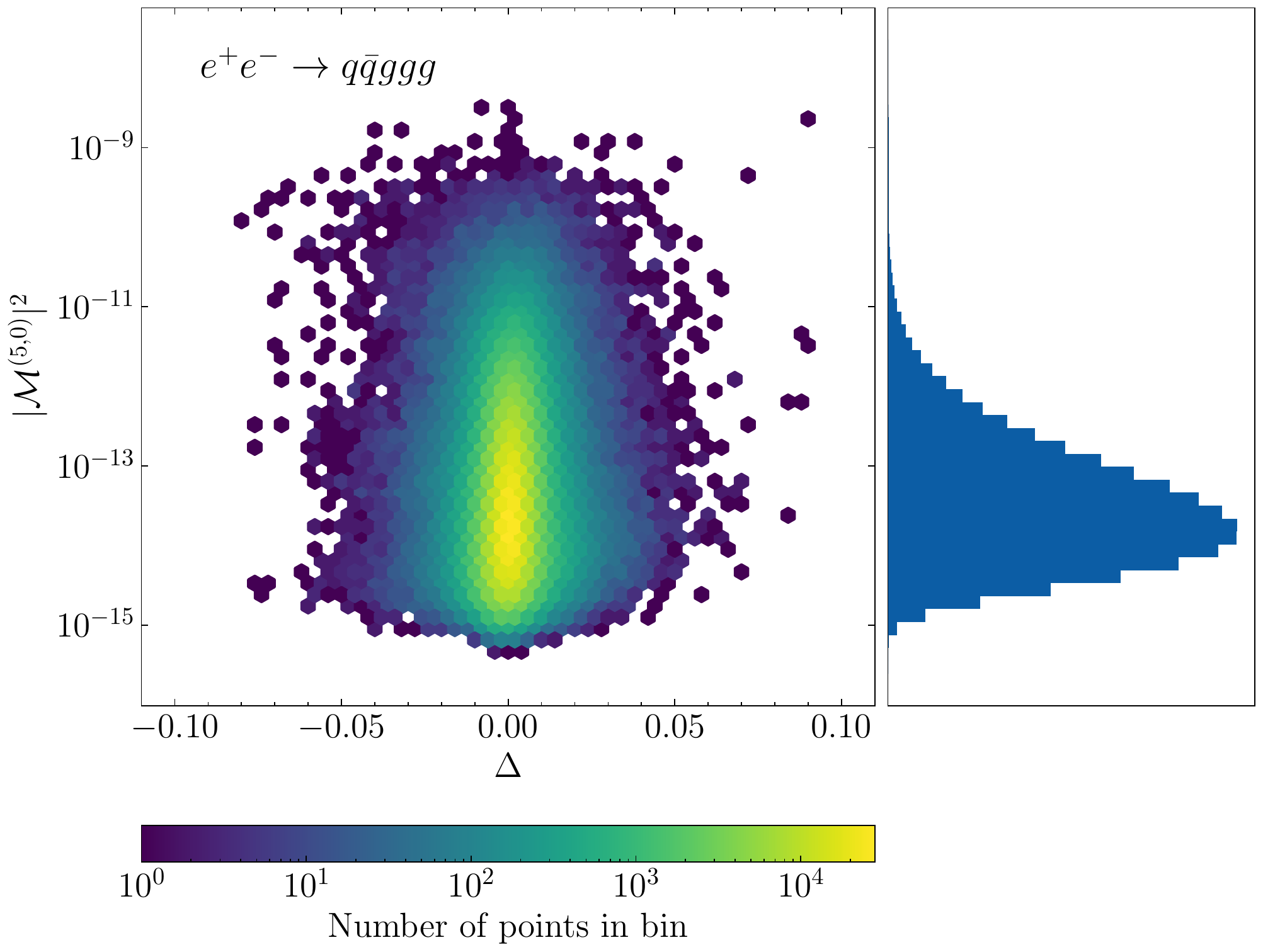}
    \caption{Left: $\Delta$ error distribution plotted against the
    tree-level matrix element. Yellow bins indicates high density regions
    of points and purple bins indicates single points.
    Right: marginal distribution of tree-level matrix elements.
    This illustrates that the network is able to reproduce
    a good approximation across all sampled phase-space.}
    \label{fig:delta_hexplot}
\end{figure}

To show that we are accurate across the entire span of the tree-level
matrix elements, and to have a closer inspection of the tails
of the $\Delta$ distribution, we plot a 2d histogram of $\Delta$
against the value of $|\mathcal{M}^{(5, 0)}|^{2}$ in
Figure~\ref{fig:delta_hexplot}. The bulk of phase-space points
are contained in the high population bins depicted in yellow,
representing the peaked distribution seen in Figure~\ref{fig:delta_distribution},
whereas the green to purple coloured bins represent the tails of the
low population $\Delta$ distribution. We see that the accuracy
stays contained inside a band and does not flare out
as the magnitude of the tree-level matrix element increases.
This shows that we manage to fit the k-factor even in
the infrared and collinear limits where the tree-level
matrix element becomes large. On the right-hand side subplot we plot
the distribution of the tree-level matrix elements where
we can see that that even with relatively few training
points in the tails, the emulator is still able to predict
these regions as well as where there is more abundant data.

\begin{figure}
    \centering
    \includegraphics[width=\linewidth]{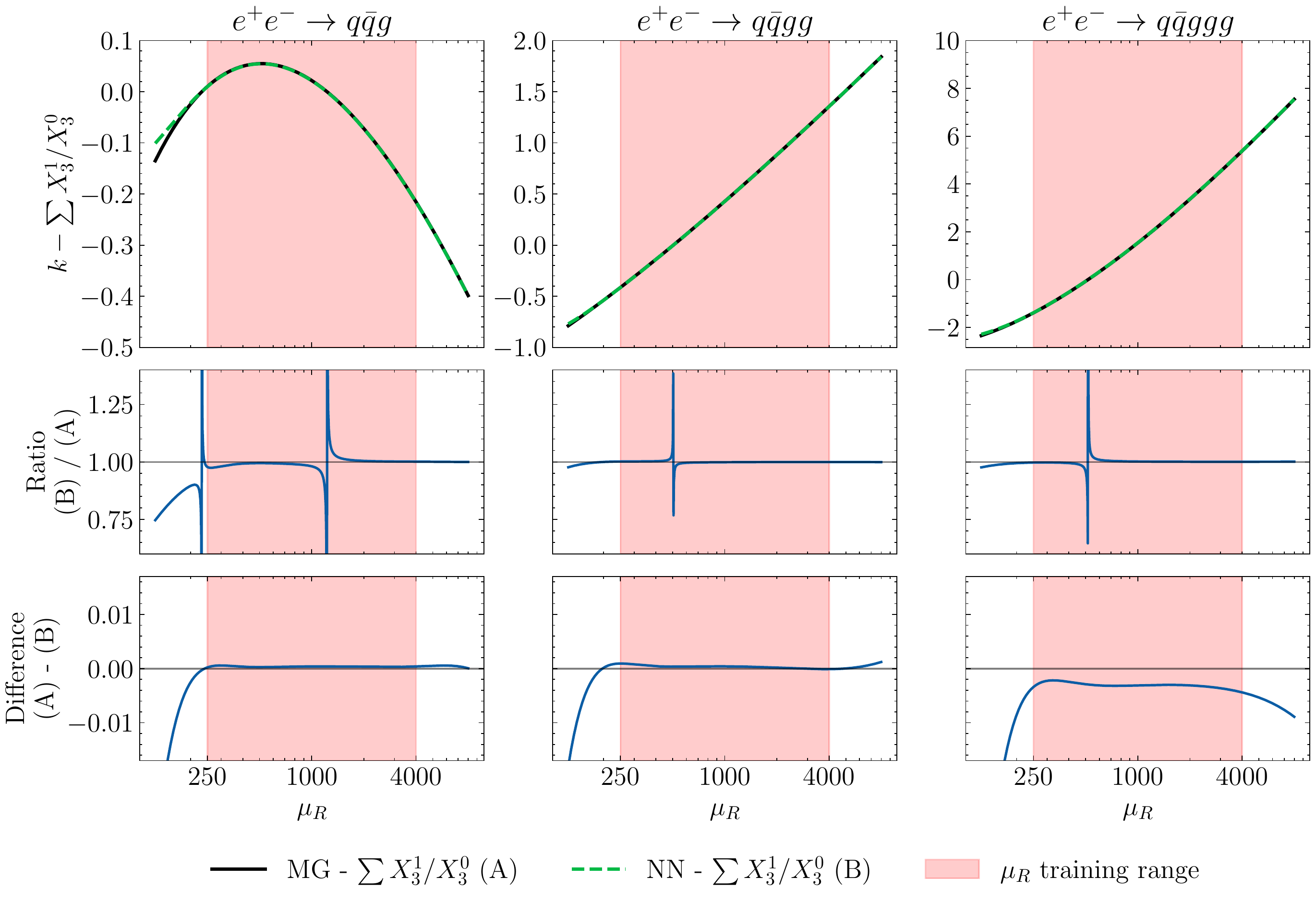}
    \caption{Renormalisation scale trajectories for all multiplicities.
    These trajectories are predictions of the k-factor with the sum
    of antenna functions subtracted to show renormalisation scale
    dependence in the remainder. Each trajectory is at one phase-space point
    sampled from {\RAMBO}, with $\mu_{R}$ spanning the range
    $[\sqrt{s_{\mathrm{com}}}/8, 8 \sqrt{s_{\mathrm{com}}}]$. Each trajectory
    is composed of 1000 points.
    The region that $\mu_{R}$ was uniformly sampled from
    for training is indicated as a red band. The error bands on the NN
    predictions are too small to be seen.}
    \label{fig:mu_trajectory}
\end{figure}

In Figure~\ref{fig:mu_trajectory}, we show that our emulator
has learned the renormalisation scale dependence, independent
of the antenna functions. To produce
a trajectory we first sample a phase-space point with
the same cuts as described in Section~\ref{sec:nn}, then we
evaluate the k-factor at this phase-space point with $\mu_{R}$
varying from $\sqrt{s_{\mathrm{com}}}/8$ to $8\sqrt{s_{\mathrm{com}}}$.
We choose to sample from a wider range than used for training
to examine the $\mu_{R}$ extrapolation performance.
After subtracting the sum of the antenna functions we see that the
remainder still has a dependence on the renormalisation
scale that is accurately captured by the neural network.
As with the $\Delta$ distribution plots in Figure~\ref{fig:delta_distribution},
there is a slight decrease in accuracy as we increase the
multiplicity, however, the ratios and differences are well-behaved
throughout the entirety of the trajectories inside the range of
training data. The only anomaly occurs when the trajectories
cross zero, causing spikes in the ratios, but the difference in
truth and prediction remains well-behaved around these regions.
Outside of the training range we see an acceptable extrapolation,
but given that the training range is wider than the range in
which the renormalisation scale is normally varied for scale
variations, we do not find this problematic.

\begin{figure}
    \centering
    \includegraphics[scale=0.5]{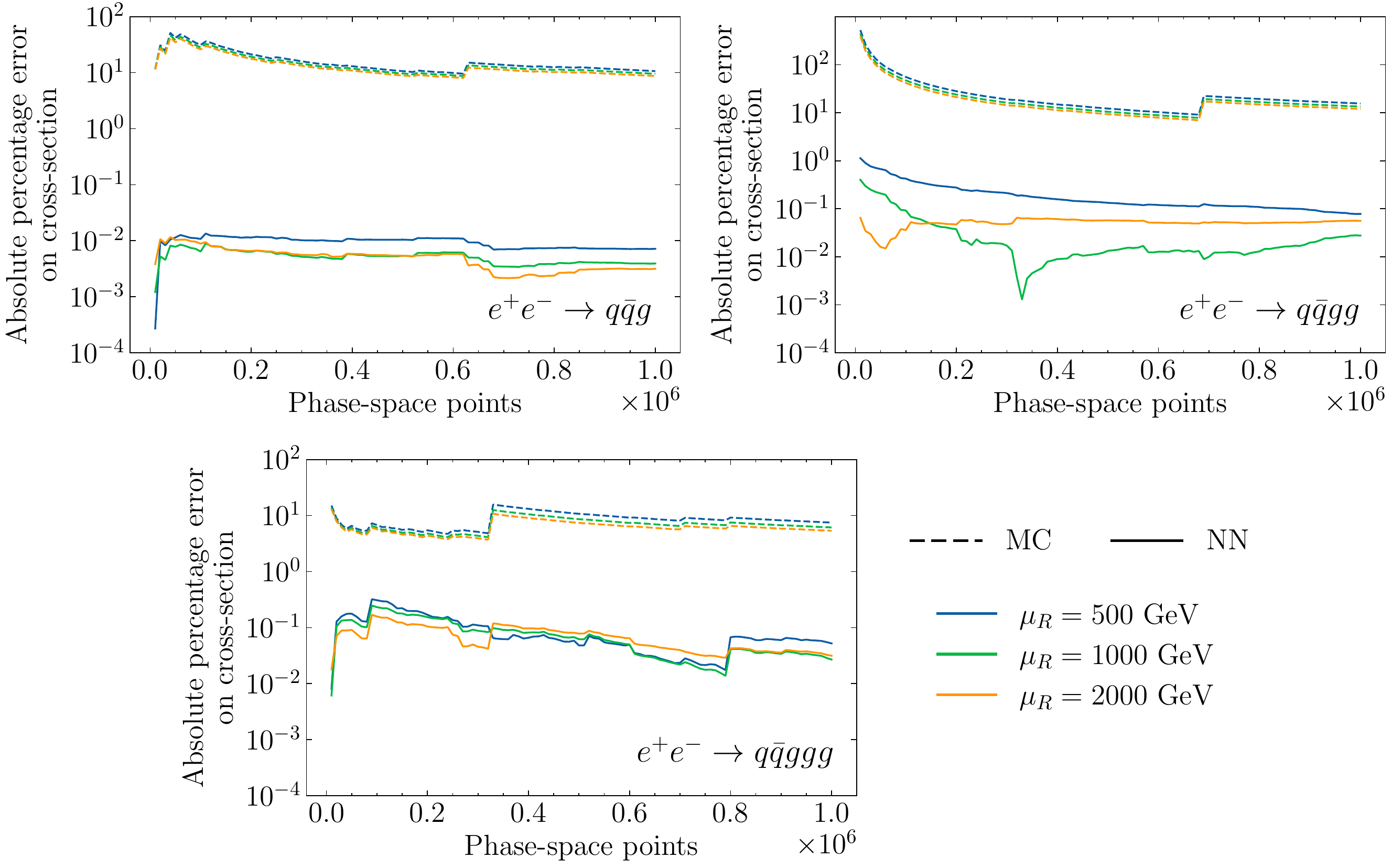}
    \caption{Error on the total cross-section across an integration
    of 1M phase-space points. For each multiplicity, we
    evaluate the matrix elements at the three values of the 
    renormalisation scale as reported.
    The solid lines are the absolute percentage error in the true total
    cross-section and the NN predicted value. The dashed
    line represents the statistical Monte Carlo integration error
    which falls as $1/\sqrt{N}$. The jumps in error are due to large
    values of the matrix element being integrated.}
    \label{fig:cross_section}
\end{figure}

\begin{figure}
    \centering
    \includegraphics[scale=0.5]{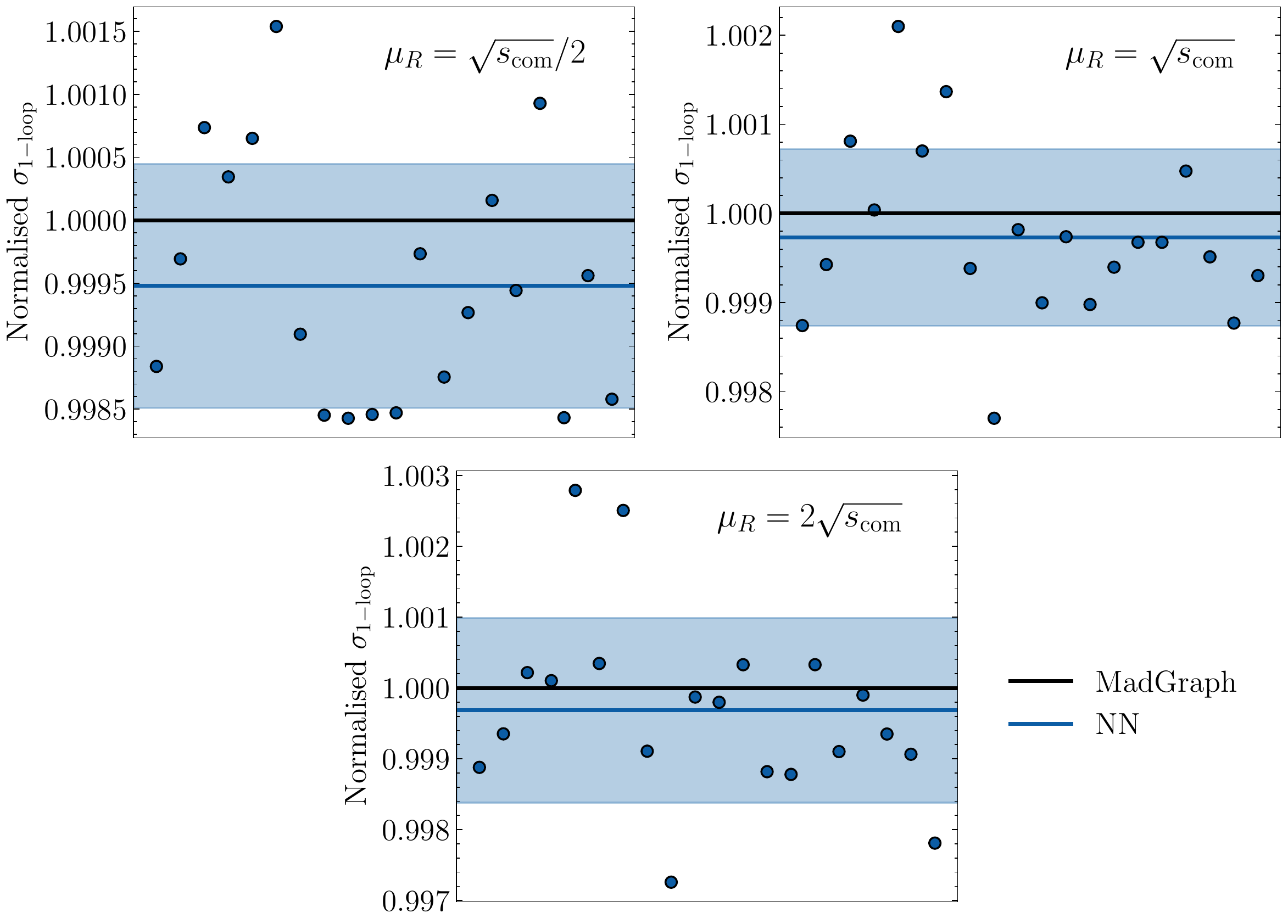}
    \caption{Scatter plot showing the variation in predictions of
    the 5 jet total cross-section for the 20 replicas.
    Total cross-sections are evaluated at the
    renormalisation scale quoted, and normalised by
    the {\MadGraph} prediction (horizontal black line).
    The mean prediction of the replicas is drawn as the horizontal
    blue line with one standard deviation illustrated by the blue band.}
    \label{fig:5j_cs}
\end{figure}

In Figure~\ref{fig:cross_section} we show that we reproduce
the total cross-sections over an integration of 1M phase-space
points, for all multiplicities, at three
different values of the renormalisation scale representing the
nominal value ($\mu_{R} = \sqrt{s_{\mathrm{com}}} = 1000$ GeV)
and the two variations usually used to estimate scale uncertainties.
Note that these phase-space points were
generated independently of those in Figure~\ref{fig:delta_distribution},
and that we are integrating the tree-level and one-loop interference, not the k-factor.
We multiply our NLO k-factor prediction with the {\MadGraph} tree-level
matrix element to reproduce the loop-matrix element for integration.
The neural network errors are well
below the statistical Monte Carlo integration error. By
neural network error we are referring to the absolute percentage
difference to the cross-section, and not the errors due to neural
network optimisation. For that, we examine the variations in the 20
replica model predictions in Figure~\ref{fig:5j_cs} where we plot the
total cross-section predictions as a scatter plot. The blue band
illustrated is one standard deviation of the 20 predictions made.
We see that the true value of the total cross-section is within
the one standard deviation band. Not shown in the figure is the
Monte Carlo statistical error which as seen in Figure~\ref{fig:cross_section}
dominates the absolute error between the NN predictions and
the true value.


Since the discrepancy between the true total
cross-section and NN predicted total cross-section is so small,
this strongly indicates that once we fit on the relatively small
training set, we can predict with good confidence on many more
phase-space points to get a prediction of the cross-section before
reaching the same level of error as the Monte Carlo statistical error.
This can also be seen from the shape of the NN error, it is relatively
constant once enough points have been integrated. This feature 
enables us to use our trained model to augment the dataset 
size by orders of magnitude beyond the size of the training set used to fit it 
and reduced the statistical error without compromising accuracy.  

\begin{figure}
    \includegraphics[width=\linewidth]{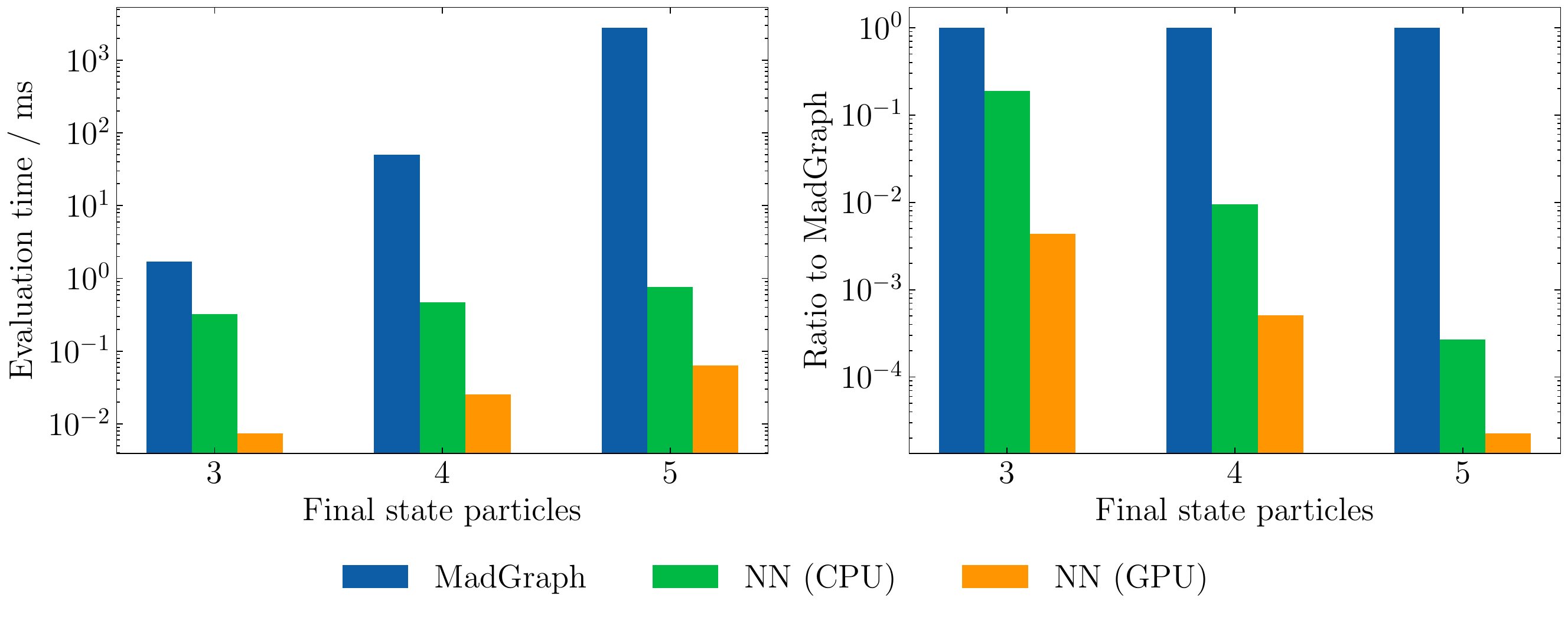}
    \caption{Left: evaluation time in milliseconds of a single phase-space point.
    Times quoted for {\MadGraph} are averaged over 1000 random phase-space
    points. Times quoted for NN are averaged over 1M random phase-space
    points. Right: ratio of evaluation times to {\MadGraph}. GPU used is
    Nvidia P100 16GB, and CPU is Intel(R) Xeon(R) Gold 6126 CPU @ 2.60GHz.}
    \label{fig:speed}
\end{figure}

In Figure~\ref{fig:speed}, we plot the evaluation time of the emulator
for both the CPU and {\CUDA} (GPU) execution providers in the {\ONNX} runtime,
as well as the reference time from {\MadGraph}. For the NN (GPU)
predictions we predict on 1M phase-space points concurrently, whereas
for the NN (CPU) predictions, we predict on one point at a time.
The times reported are then the mean of the total evaluation times.
We see that compared to {\MadGraph}, our NN emulator
is faster for all multiplicities, although the advantage is greatest
for the 5 jet case, with speed gains of over four orders of magnitude
when utilising GPU acceleration. The advantage of using the GPU is
not only from being able to batch process the predictions, it is
also to leverage the auto-vectorisation tools provided
by, for example {\TensorFlow}, to accelerate the model input computations.

While the evaluation time of the GPU accelerated NNs are by far
the quickest, and would be the ideal scenario for a NN emulator to be used,
event generation typically occurs on CPUs where phase-space
points are evaluated one at a time. This is precisely
why our NN (CPU) predictions were made on single phase-points and
not over batches, to showcase what the performance would be like
when embedded in a typical workflow. We observe that even with
this constraint, the NN emulator is much quicker in the higher
multiplicity cases.

\begin{figure}
    \centering
    \includegraphics[width=\linewidth]{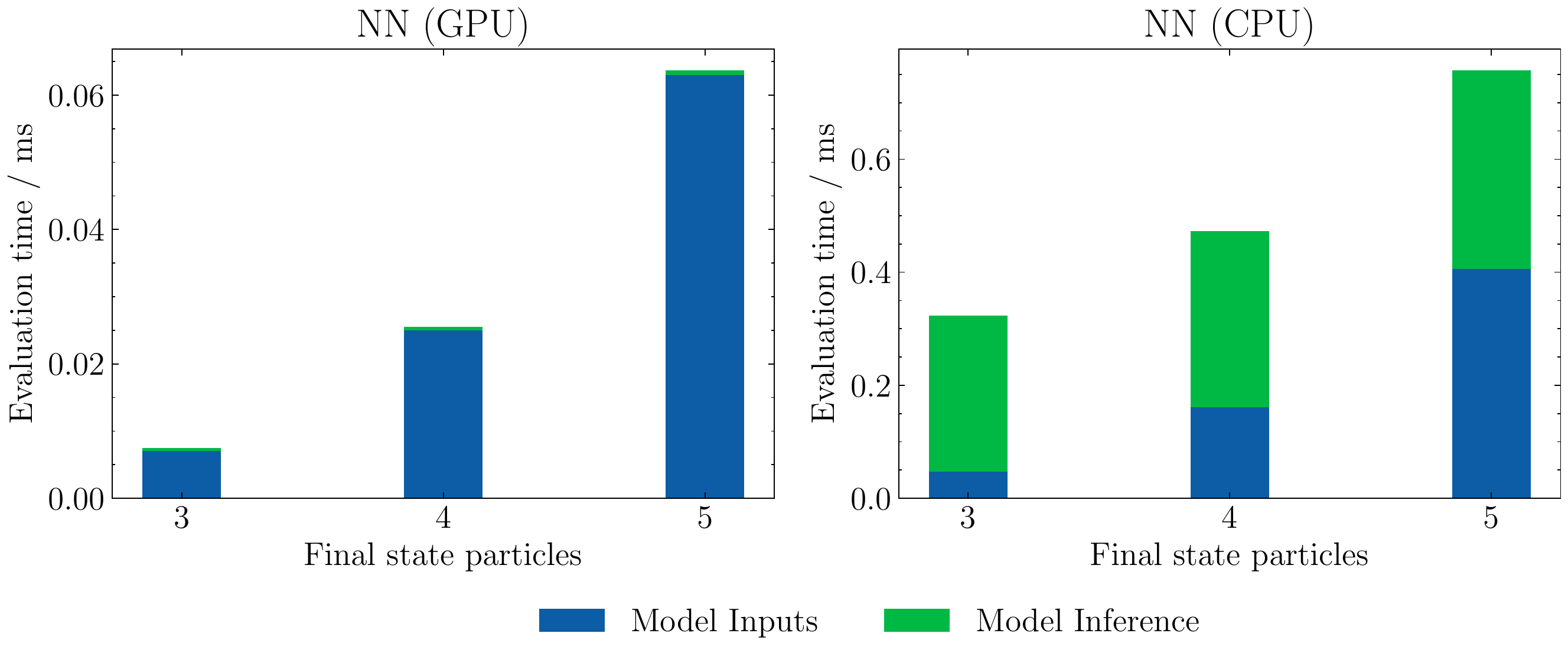}
    \caption{Breakdown of the total time taken to predict on a phase-space
    point (averaged over 1M predictions) into model input
    computations and the actual model inference for both GPU and CPU
    model deployments. The model inference
    portions in the NN (GPU) subplot are very small.}
    \label{fig:speed_breakdown}
\end{figure}

One of the main bottlenecks in the NN (CPU) prediction is using
the NN ensemble to infer on single phase-space points, this is
illustrated by the weak scaling in multiplicity
in the left subplot of Figure~\ref{fig:speed} and is illustrated
explicitly in Figure~\ref{fig:speed_breakdown}. Since the model
architecture is identical for all multiplicities other than the
final output layer, there will not be much difference in cost.
Another bottleneck is the computation of model inputs, which contain
large, complex expressions with many evaluations of logarithms and
dilogarithms. In our $\mathtt{Python}$ implementation\footnote{
    We anticipate that a \texttt{C++} implementation of
    the input computations would be significantly more performant,
    however, the {\ONNX} model prediction is already highly
    optimised.
},
the computation time of these inputs is comparable to the
model inference time in the 5 jet case, as shown in Figure~\ref{fig:speed_breakdown}.
Time taken to compute the model input scales with the final-state
partons because of the increase in number of antenna functions in the
ansatz to account for the large number of singular configurations,
as well as a larger number of the other input variables.

For the NN (GPU) predictions, the model inference time is negligible
compared to the model input computations since we take advantage
of predicting on the entire batch of 1M phase-space points at once,
and so once averaged across this dataset each single point takes
an insignificant amount of time.
The model input computations are vectorised on the GPU and so we see
an order of magnitude reduction in time taken to calculate them compared
to computing them on a single core of a CPU.

We note that our emulator retains good performance even
when we go to higher multiplicity. In Ref.~\cite{Badger:2022hwf}
the authors 
expressed concern about the fact that the accuracy of the surrogate model
is decreasing with the multiplicity of the process.
We show that by incorporating suitable physically motivated
functions in the ansatz that the network accuracy drops off much less
rapidly when going to higher multiplicities.

\section{Conclusion}\label{sec:conclusion}
In this article we presented the extension to one-loop matrix elements of Ref.~\cite{Maitre:2021uaa},
where we introduced the factorisation-aware model for tree-level processes. By adapting the
ingredients provided to the neural network model to a suitable
set, we have been able to adapt the emulation of tree-level
matrix element to the modelling of NLO k-factors. We have shown that the philosophy of
incorporating relevant physics information into a regression model greatly
improves accuracy of predictions even for matrix elements with a
 complex divergence structure.

The results presented demonstrate that predictions are at the percent
level for the most demanding process, with accuracy in the infrared
regions of phase-space being well-behaved. We have also shown that
for relatively few training points, the model is able to learn the
target function well enough such that the accumulated error due to the
modelling is well below the statistical integration uncertainty of the
training sample.    
Furthermore, the modelling uncertainty of cross-section predictions
associated with optimisation of model parameters was shown to
contain the truth value and therefore provides a useable estimate of 
the accuracy of the emulation. 

In addition to being accurate, we have given evidence that the
model, although optimally deployed on a GPU, is orders of magnitude
quicker than traditional loop providers on a single CPU. By deploying
the model with the {\ONNX} runtime a generic interface to the
neural network model is available in many programming languages,
allowing it to be embedded into modern event generators which are
mainly written in \texttt{C++}.

These facts put together provide conclusive evidence that the factorisation aware neural
network model can be used to augment existing samples with
additional phase-space points with confidence. Given the high compuational 
costs of high-multiplicity one-loop matrix elements using this data augmentation
capability can further the reach of existing and future simulations for a 
fixed computational resource envelope.

\acknowledgments
We would like to thank Oscar Braun-White for useful discussions
on antenna functions and for providing code for cross-checking
our implementation of them. We would also like to thank Timo
Janßen for introducing us to the {\ONNX} runtime.
\hspace{2cm}
\hrule

\appendix
\section{Finite part of singularity operators}\label{appendix:finite_terms}
The finite part of the Catani singularity operators can be written as
\begin{equation}
    \mathcal{F}(\mathbf{I}^{(1)}_{ij}(\epsilon, s_{ij})) = \epsilon_{0} + \epsilon_{1} \Re{(z)} + \dfrac{1}{2}\epsilon_{2} \Re{(z^{2})} \, ,
\end{equation}
where
\begin{equation}
    z = \zlog(\mu_{R}^{2}) - \zlog(-s_{ij})
\end{equation}
and coefficients are given as
\begin{align}
    \epsilon_{0} &= \dfrac{\pi^{2}}{24} \, , \nonumber \\
    \epsilon_{1} &=
    \begin{cases}
        -\frac{5}{6}, & \text{if } ij = qg \; \text{or} \; gq \\
        -\frac{3}{4}, & \text{if } ij = qq \\
        -\frac{11}{12}, & \text{if } ij = gg 
    \end{cases} \\
    \epsilon_{2} &= -\dfrac{1}{2} \, . \nonumber
\end{align}
The $\zlog$ function is extending the logarithm for all real-values
\begin{align}
    \zlog(x) = 
    \begin{cases}
        \log(x), & \text{if } x \geq 0 \\
        \log(|x|) - i \pi , & \text{otherwise} \, .
    \end{cases}
\end{align}

\section{Antenna functions}\label{appendix:antenna_functions}
Here we tabulate the full set of antenna functions
which we build into our emulation model. In the main article we
refer to the antenna functions as $X_{ijk}^{\ell}$ which
in practice is replaced with specific antennae containing either
$qg\bar{q}$, $qgg$ ($\bar{q}gg$), or $ggg$, which are
referred to as $A$, $D$, and $F$, respectively. The antennae
listed in Table~\ref{table:antenna_functions} are sufficient to
describe all infrared singularities in the partonic
processes we consider.
\begin{table}
    \centering
    \begin{tabular}{cccc}
        \toprule
        $n$ & Tree-level antenna & One-loop antenna & {\{$i$, $j$, $k$\}} permutations \\
        \midrule
        3 $(q \bar{q} g)$     & $A_{3}^{0}(q, g, \bar{q})$ & $A_{3}^{1}(q, g, \bar{q})$  & (1, 3, 2) \\
        \midrule
        4 $(q \bar{q} g g)$   & \makecell{$A_{3}^{0}(q, g, \bar{q})$ \\ $D_{3}^{0}(q, g, g)$} & \makecell{$A_{3}^{1}(q, g, \bar{q})$ \\ $D_{3}^{1}(q, g, g)$} & \makecell{(1, 3, 2), (1, 4, 2) \\ (1, 3, 4), (2, 3, 4)} \\
        \midrule
        5 $(q \bar{q} g g g)$ & \makecell{$A_{3}^{0}(q, g, \bar{q})$ \\ $D_{3}^{0}(q, g, g)$ \\ $F_{3}^{0}(g, g, g)$} & \makecell{$A_{3}^{1}(q, g, \bar{q})$ \\ $D_{3}^{1}(q, g, g)$ \\ $F_{3}^{1}(g, g, g)$} & \makecell{(1, 3, 2), (1, 4, 2), (1, 5, 2) \\ (1, 3, 4), (1, 3, 5), (1, 4, 5), \\ (2, 3, 4), (2, 3, 5), (2, 4, 5), \\ (3, 4, 5)} \\
        \bottomrule
    \end{tabular}
    \caption{List of antenna functions we use for each process,
    and the full list of {\{$i$, $j$, $k$\}} permutations, where $q=1$,
    $\bar{q}=2$, $g=3,4,5$.}
    \label{table:antenna_functions}
\end{table}

\bibliographystyle{JHEP}
\bibliography{nn_me_1L}

\end{document}